%% file: main.tex
\documentclass[a4paper,11pt]{article}
\usepackage{jinstpub} 
\usepackage{lineno}
\usepackage{soul}
\usepackage{xcolor}

\title{\boldmath  VUV Reflectance Measurements for Materials Relevant to Argon and Xenon Experiments }

 \author[a,b,c,1,2]{J. Soto-Oton,\note{Corresponding author.}\note{Work performed while at IFIC-CSIC/UV, Spain (current address: Nikhef)}}
 \author[c]{H. Amar,}
 \author[c]{A. Cervera,}
 \author[c]{A. Roche}
 \affiliation[a]{Universiteit van Amsterdam, NL-1098 XG Amsterdam, The Netherlands}
 \affiliation[b]{Nikhef National Institute of Subatomic Physics, 1098 XG Amsterdam, The Netherlands}
 \affiliation[c]{Instituto de Física Corpuscular, CSIC and Universitat de València, 46980 Paterna, Valencia, Spain}

\emailAdd{j.soto@cern.ch}

\abstract{
Accurate knowledge of material reflectance in the vacuum ultraviolet (VUV) range is crucial for optimizing photon detection in noble gas detectors such as DUNE. Despite its importance, reflectance values for detector materials in the VUV region remain poorly characterized, with literature values showing significant variation depending on surface termination and finish. An angular-resolved reflectance measurement system developed at the Instituto de Física Corpuscular at Valencia (IFIC) that operates in a gaseous argon atmosphere is presented, enabling realistic measurements of detector materials under controlled conditions. The setup couples a deuterium lamp to a monochromator and employs a motorized PMT rotating around the sample to measure reflected light distributions across a wide angular range. We have characterized two key DUNE materials—aluminum field cage profiles and stainless steel cryostat membranes—in both the UV-VIS (300–500 nm) and VUV (128–200 nm) ranges. In the UV-VIS region, we confirm literature values of approximately 60\% reflectance for aluminum and 40\% for stainless steel. Preliminary VUV measurements at 45° angle of incidence yield reflectance values of 10–15\% for both materials, significantly lower than their UV-VIS counterparts. The reflected light distributions exhibit a mixed character between specular and diffuse reflection. These results have direct implications for detector simulations and light yield predictions in next-generation experiments.

}


\arxivnumber{XXXXX} 

\begin{document}
\maketitle
\flushbottom

\section{Introduction}
\label{sec:intro}
\input{1_Introduction}

\section{Experimental setup}
\label{sec:setup}
\input{2_Setup}    

\section{Methodology}
\label{sec:methods}
\input{3_Methodology}

\section{Results}
\label{sec:results}
\input{4_Results}    

\section{Discussion}
\label{sec:discussion}
\input{5_Discussion}

\section{Conclusions and future steps}
\label{sec:conclusions}
\input{6_Conclusions}

\acknowledgments
The authors acknowledge support from Agencia Estatal de Investigación (projects PID2023-147949NB-C52 and JDC2023-051298-I) and Generalitat Valenciana (ASFAE/2022/029 and ASFAE/2022/028).



\bibliographystyle{JHEP}
\bibliography{biblio.bib}

\end{document}

%% file: 1_Introduction.tex
Scintillation light in noble element detectors serves multiple critical functions: it provides a trigger signal for the liquid-argon time projection chambers as in the Deep Underground Neutrino Experiment (DUNE) \cite{DUNE:2020lwj}, enables precise timing reconstruction ($t_{0}$), and allows for calorimetric reconstruction of deposited energy. Reliable calorimetric reconstruction depends critically on accurate knowledge of the light yield through the active volume, defined as the number of photons detected per unit deposited energy (typically expressed in photons/MeV). To achieve this accuracy, a comprehensive understanding of photon propagation within the detector is essential.

A key aspect often overlooked in detector design and simulation is the role of material reflectance. The surfaces of the main detector components can reflect a significant fraction of scintillation photons back into the active volume. Poor characterization of reflectance properties can introduce a significant bias in the light yield determination \cite{10.1117/12.934138}. In particular, it has been demonstrated in \cite{SotoOton:2022pgz} that considering a reflectance of 40\% for the liquid-argon scintillation light on the aluminium field cage and stainless steel cryostat walls of ProtoDUNE Dual-Phase \cite{DUNE:2022ctp} would lead to a bias of 40\% in the determination of the simulated light yield with respect to the null reflectance case. Despite this sensitivity, reflectance data for detector materials in the vacuum ultraviolet (VUV) range—where argon and xenon scintillation peaks occur—remain poorly characterized in the literature, with published values showing substantial variation depending on surface termination and finish.

To address this critical gap, we have developed a dedicated angular-resolved reflectance measurement system at IFIC, operating in a gaseous argon (GAr) atmosphere to enable realistic optical characterization of DUNE prototype materials under controlled conditions. Additionally, an angular-resolved measurement would improve our understanding of the photon propagation inside the detector, and allow the refinement of the simulation by implementing a more realistic angular distribution. This work provides the first systematic measurements of aluminum and stainless steel reflectance in the VUV range with angular resolution.

%% file: 2_Setup.tex
The measurement system couples a deuterium lamp (model McPherson 634 \cite{Mcpherson:DeLamp})  and a tungsten lamp  (model McPherson 621 \cite{Mcpherson:TgLamp}) to a McPherson 302/234 monochromator \cite{Mcpherson:MonoChromator}. This device incorporates an aberration-corrected concave holographic diffraction grating of 1200 grooves/mm, in a Seya-Namioka optical design. The grating is equipped with a wavelength-optimized reflective coating of aluminium, enhanced with magnesium fluoride. This enables a wavelength selection range of 115-550 nm. The monochromator output is coupled to a collimator and a methacrylate black box where the measurement is performed. When measuring in the VUV range, the entire system is maintained in a pure GAr atmosphere (Air Liquid ALPHAGAZ~\cite{Ar} of 99.999\% purity) within a dedicated dark room to allow for measurements in the VUV region. The black box is continuously purged with pure GAr and kept at a slight overpressure of about 100 mbar, monitored by a pressure sensor. This atmosphere choice unlike traditional vacuum measurements serves two key purposes: (i) it provides the optical transmission necessary for VUV measurements while enabling flexibility to fully instrument the black box and (ii) it protects the monochromator grating from contamination, as outgassing in vacuum-based VUV systems can lead to the deposition of carbonaceous films on optical elements and degrade their performance over time~\cite{10.1116/1.2140005,DOLGOV2015708}. 

\begin{figure}[htbp]
\centering
\includegraphics[width=.99\textwidth]{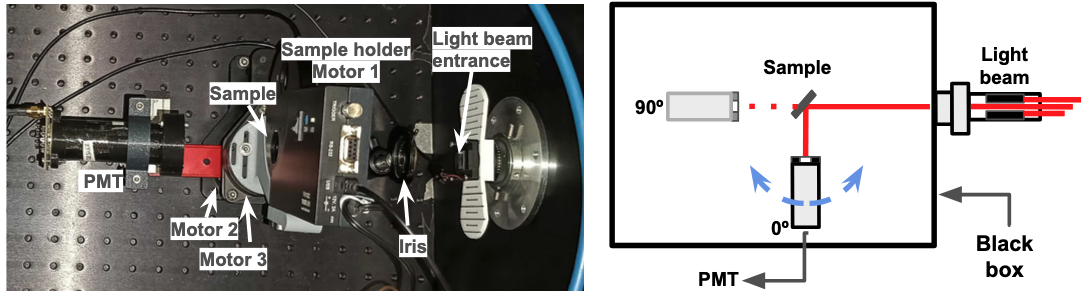}
\caption{A picture of the measurement setup inside the black box is shown on the left, while a diagram of the setup on the right.\label{fig:setup}}
\end{figure}

The angular-resolved measurement capability is provided by a 3D-printed assembly comprising three automated motorized stages. Motor 1 allows sample exchange, Motor 2 rotates a VUV-sensitive PMT (Hamamatsu R6836-Y004) around the sample to set the angle of reflection, and Motor 3 rotates the sample holder to set the angle of incidence (AOI). A rectangular mask ($2\times20$~mm$^{2}$ opening providing 1.6\textdegree~angular resolution) placed in front of the PMT detector enhances angular resolution and defines the collection solid angle. A picture and a diagram of the setup are shown in Fig.~\ref{fig:setup}. Since the monochromator provides a high light intensity, the PMT is operated without gain by measuring the first-dynode current with a picoammeter. A 100 V bias between the photocathode and the first dynode ensures full collection efficiency.

Two materials from the DUNE prototypes are measured: (i) 6101 aluminium sample (Al) extracted from a field cage profile and (ii) 304L stainless steel sample (SS) from the cryostat inner membrane \cite{DUNE:2021hwx}. Calibration samples—including a UV mirror and a UV beam splitter with known reference reflectance values—were employed to validate the measurement methodology across the UV-VIS range.

%% file: 3_Methodology.tex
For each wavelength and at a fixed angle of incidence, the angular distribution of reflected light is recorded by rotating the PMT across a defined angular range (typically $\pm 40^\circ$ from the specular direction. Direct-light normalization measurements (when the PMT is around $90^{\circ}$ in Fig.~\ref{fig:setup}) are acquired at each wavelength both before and after sample measurements to account for variations in the deuterium lamp intensity and GAr atmosphere transmission stability (typically maintained to less than 6\% in all measurements). The uncertainty introduced by this variation is added in quadrature for each measurement.

\begin{figure}[htbp]
\centering
\includegraphics[width=.45\textwidth]{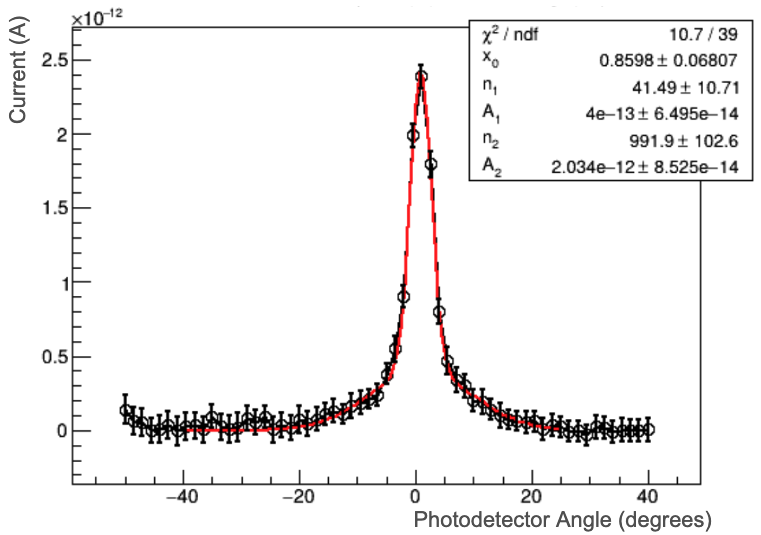}
\qquad
\includegraphics[width=.45\textwidth]{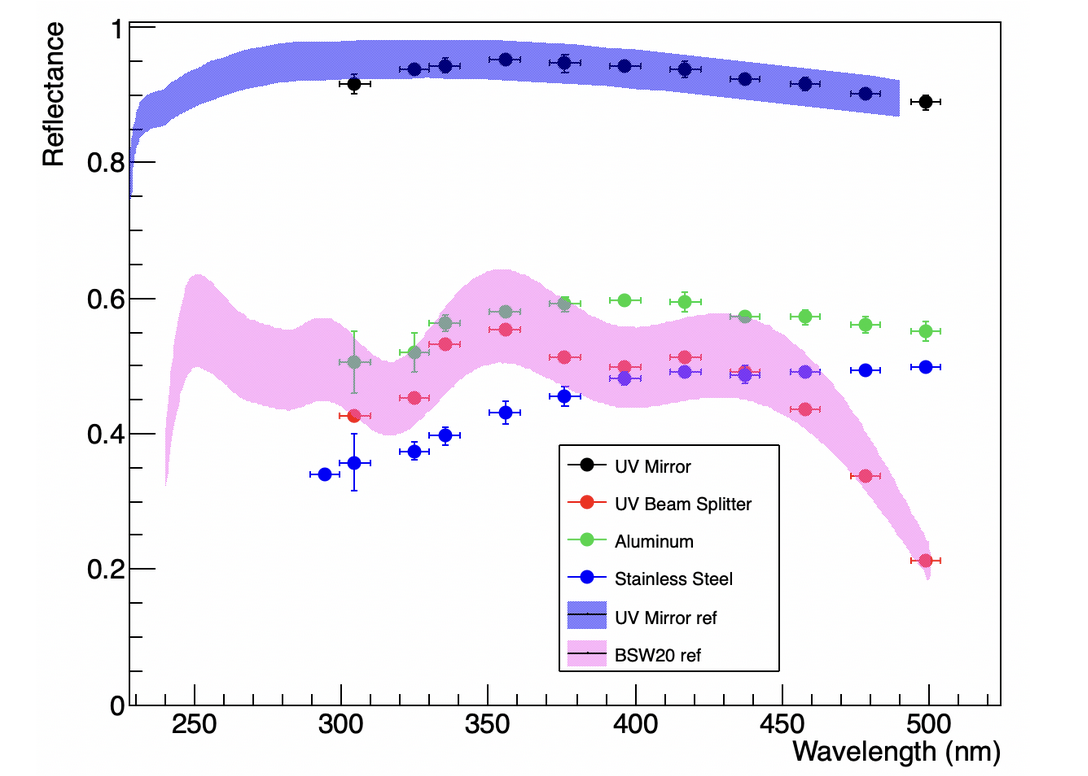}
\caption{Left: Angular-resolved reflectance intensity for the stainless-steel sample at an angle of incidence of $45^{\circ}$ and a wavelength of 300~nm in air\label{fig:phong}. Right: Reflectance of the SS and Al samples in the UV/Visible range in air.\label{fig:visref}}
\end{figure}

The reflected light distributions exhibit a mixed character between specular and diffuse reflection, well described by the following equation, inspired by the shading-function model described in \cite{Phong}: $I(\theta)=\sum_{i=1}^{2} A_{i}cos^{n_{i}} (\theta - x_{0})$, 
where $n$ parameterizes surface roughness (higher $n$ indicates more specular behavior), $x_0$ is the specular direction, $A$ is the normalization amplitude, and $i$ is the summation index. This functional form successfully captures the beam profiles observed in our data. An example fit to data with the resulting parameter values is shown in the left panel of Fig.~\ref{fig:phong}. In general SS shows a more specular behavior than Al ($n$ parameter: \textasciitilde{}1000 vs. \textasciitilde{}55 at 300-500~nm). 

The integrated (or hemispherical) reflectance is computed by summing the angular-resolved measurements (when the PMT is around $0^{\circ}$ in Fig.~\ref{fig:setup}) and normalizing to direct light measurement (when the PMT is around $90^{\circ}$ in Fig.~\ref{fig:setup}): $R = \sum_{\theta=-40^{\circ}}^{40^{\circ}} I_\theta / \sum_{\theta=80^{\circ}}^{100^{\circ}} I_\theta$.
For metallic samples (Al and SS), which exhibit more diffuse reflection, the azimuthal scan covers the full range of interest ($\pm40^{\circ}$) with PMT rotation, while vertical acceptance is limited by the mask height (2 cm or $\pm$8$^{\circ}$). Radial symmetry is assumed to estimate the uncollected light in the vertical direction.

%% file: 4_Results.tex
\subsection{UV-Visible Measurements (300–500 nm) in air}

The right panel of Fig.~\ref{fig:visref} shows the reflectance measurements in the UV-VIS range in air. They show excellent agreement with the expected values for the reference samples: the UV mirror (Thorlabs PF1011-F01~\cite{Thorlabs:PF1011}) and beamsplitter (Thorlabs BSW20~\cite{Thorlabs:BSW20}) reproduce their manufacturer specifications. For DUNE materials, aluminum exhibits approximately 60\% reflectance across the UV-VIS range, while stainless steel shows approximately 40\%, consistent with literature values, but in this case, for real detector samples \cite{10.1117/12.934138,NAMBA1980409, Madden:63, Yang:06}.


\subsection{VUV Measurements (128–200 nm) in GAr}

\begin{figure}[htbp]
\centering
\includegraphics[width=.99\textwidth]{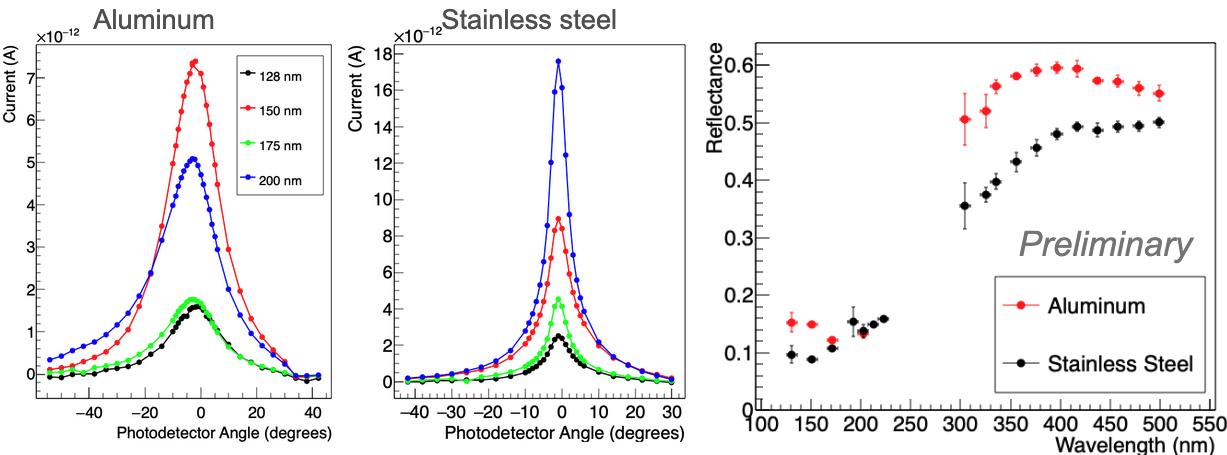}
\caption{Left: Angular-resolved reflected intensity of the SS and Al samples at VUV wavelengths. Right: Reflectance of the SS and Al samples in the VUV and UV/Visible range.\label{fig:vuvspectrums}}
\end{figure}

Despite the reduced photon flux and lower detector sensitivity in the VUV, reflected light is successfully detected from both materials, as shown in Fig.~\ref{fig:vuvspectrums}. The angular distributions at VUV wavelengths (128–200 nm) remain qualitatively similar, with SS sample showing more specular behavior than Al sample considered in these tests (higher $n$ parameter: \textasciitilde{}600 vs. \textasciitilde{}80 at 128~nm), and with small wavelength dependence.

\subsection{Integrated reflectance}

At an angle of incidence of 45$^{\circ}$ in the VUV range, preliminary reflectance values are 10\% (15\%) for SS (Al) samples, significantly lower than their UV-VIS counterparts, as it is shown in the right panel of Fig.~\ref{fig:vuvspectrums}. Al sample is slightly more reflective than SS sample, but far below some literature values \cite{Yang:06}, highlighting the strong dependence on surface finish and the need for sample-specific measurements. This substantial drop at shorter wavelengths reflects the wavelength-dependent optical properties of these materials and has important consequences for detector design.


%% file: 5_Discussion.tex
A significant achievement of this work is the successful operation of an optical measurement setup capable of performing angular-resolved reflectance measurements in the VUV range using a gaseous argon atmosphere. Unlike vacuum-based systems, this approach provides substantial experimental flexibility: motorized stages, electronic components, and complex instrumental geometries can be readily deployed without the constraints imposed by vacuum compatibility. This versatility enables sophisticated measurements that would be significantly more challenging or impractical in  vacuum conditions, opening new possibilities for systematic characterization of detector materials.

The measured angular distributions reveal a mixed specular-diffuse character for both aluminium and stainless steel surfaces. This finding has important implications for photon transport modelling: simple assumptions of purely specular or purely Lambertian reflection inadequately describe the true optical behaviour. Implementation of realistic angular distributions derived from the shading-function model \cite{Phong} parametrization into Monte Carlo simulations should improve predictions of light yield.


The integrated reflectance values measured in the VUV range are lower than the reflectance values typically assumed in DUNE simulations, where about 70\% is considered for aluminium and 30-40\% for stainless steel \cite{DUNE:2023nqi}. This discrepancy implies that the contribution of field-cage and cryostat surfaces to the overall light yield should be carefully re-evaluated.


These preliminary results should be interpreted with appropriate caution. A comprehensive evaluation of measurement systematics is currently underway, with detailed findings to be presented in forthcoming publications. In particular, the extrapolation procedure used to estimate light scattered at wide angles beyond the PMT acceptance —critical for diffuse samples— requires more rigorous assessment. Refined treatments accounting for potential asymmetries and instrumental effects are under development to ensure robust determinations of the total integrated reflectance.

%% file: 6_Conclusions.tex
A dedicated facility at IFIC for angular-resolved reflectance measurements in the VUV range using a GAr atmosphere as been stablished, enabling realistic measurements of detector materials under controlled conditions. Preliminary results indicate reflectance values in the range of 10–15\% for aluminum and stainless steel at an angle of incidence of 45$^{\circ}$ in the VUV, with mixed angular distributions best described by a shading-function model \cite{Phong}. These measurements provide a foundation for improved detector simulations and will guide optimization of light collection in next-generation noble gas experiments.

Future work includes: (i) systematic studies of atmosphere stability and beam collimation effects, (ii) extension to additional angles of incidence and wavelengths, (iii) measurements under vacuum conditions, and (iv) characterization of additional materials relevant to detector design.

%% file: biblio.bib
@mastersthesis{SotoOton:2022pgz,
    author = "Soto Oton, Jose Alfonso",
    title = "{Scintillation light detection techniques in a 750-ton liquid argon TPC for the Deep Underground Neutrino Experiment}",
    reportNumber = "CERN-THESIS-2022-063",
    type = "doctoral dissertation",
    school = "2022-04-06, U. Autonoma, Madrid (main)",
    month = "4",
    year = "2022"
}

@inbook{Phong,
author = {Phong, Bui Tuong},
title = {Illumination for computer generated pictures},
year = {1998},
isbn = {158113052X},
publisher = {Association for Computing Machinery},
address = {New York, NY, USA},
url = {https://doi.org/10.1145/280811.280980},
booktitle = {Seminal Graphics: Pioneering Efforts That Shaped the Field, Volume 1},
pages = {95–101},
numpages = {7}
}

@article{DUNE:2020lwj,
    author = "Abi, Babak and others",
    collaboration = "DUNE",
    title = "{Deep Underground Neutrino Experiment (DUNE), Far Detector Technical Design Report, Volume I Introduction to DUNE}",
    eprint = "2002.02967",
    archivePrefix = "arXiv",
    primaryClass = "physics.ins-det",
    reportNumber = "FERMILAB-PUB-20-024-ND, FERMILAB-DESIGN-2020-01",
    doi = "10.1088/1748-0221/15/08/T08008",
    journal = "JINST",
    volume = "15",
    number = "08",
    pages = "T08008",
    year = "2020"
}

@article{DUNE:2022ctp,
    author = "Abed Abud, Adam and others",
    collaboration = "DUNE",
    title = "{Scintillation light detection in the 6-m drift-length ProtoDUNE Dual Phase liquid argon TPC}",
    eprint = "2203.16134",
    archivePrefix = "arXiv",
    primaryClass = "physics.ins-det",
    reportNumber = "CERN-EP-DRAFT-MISC-2022-003, FERMILAB-PUB-22-242-LBNF, CERN-EP-DRAFT-MISC-2022-003; FERMILAB-PUB-22-242-LBNF",
    doi = "10.1140/epjc/s10052-022-10549-w",
    journal = "Eur. Phys. J. C",
    volume = "82",
    number = "7",
    pages = "618",
    year = "2022"
}

@article{DUNE:2023nqi,
    author = "Abed Abud, Adam and others",
    collaboration = "DUNE",
    title = "{The DUNE Far Detector Vertical Drift Technology. Technical Design Report}",
    eprint = "2312.03130",
    archivePrefix = "arXiv",
    primaryClass = "hep-ex",
    reportNumber = "FERMILAB-TM-2813-LBNF",
    doi = "10.1088/1748-0221/19/08/T08004",
    journal = "JINST",
    volume = "19",
    number = "08",
    pages = "T08004",
    year = "2024"
}

@misc{Mcpherson:MonoChromator,
  author = {McPherson},
  title = {{200mm f.l. Vacuum Monochromator}},
  howpublished = "\url{https://www.mcphersoninc.com/pdf/234302.pdf}",
  note = "[Online; accessed 31-March-2026]"
}

@misc{Mcpherson:TgLamp,
  author = {McPherson},
  title = {{Universal Light Source System}},
  howpublished = "\url{https://mcphersoninc.com/pdf/621.pdf}",
  note = "[Online; accessed 31-March-2026]"
}

@misc{Mcpherson:DeLamp,
  author = {McPherson},
  title = {{VACUUM ULTRAVIOLET DEUTERIUM LIGHT SOURCE}},
  howpublished = "\url{https://mcphersoninc.com/pdf/634.pdf}",
  note = "[Online; accessed 31-March-2026]"
}

@misc{Thorlabs:PF1011,
  author = {Thorlabs},
  title = {{PF1011-F01}},
  howpublished = "\url{https://www.thorlabs.com/item/PF1011-F01}",
  note = "[Online; accessed 31-March-2026]"
}

@misc{Thorlabs:BSW20,
  author = {Thorlabs},
  title = {{BSW20}},
  howpublished = "\url{https://www.thorlabs.com/item/BSW20}",
  note = "[Online; accessed 31-March-2026]"
}

@misc{Ar,
  author = {Air Liquid},
  title = {{ALPHAGAZ 1 ARGON / Ar}},
  howpublished = "\url{https://es.airliquide.com/product-assets/8/a/2/5/8a2566a0a01c8645729b290b759094e99fd85550_ALPHAGAZ%E2%84%A2_1_ARG%C3%93N_es_ES_v1.3.pdf}",
  year = {2025},
  note = "[Online; accessed 31-March-2026]"
}

@article{10.1116/1.2140005,
    author = {Hollenshead, Jeromy and Klebanoff, Leonard},
    title = {Modeling radiation-induced carbon contamination of extreme ultraviolet optics},
    journal = {Journal of Vacuum Science \& Technology B: Microelectronics and Nanometer Structures Processing, Measurement, and Phenomena},
    volume = {24},
    number = {1},
    pages = {64-82},
    year = {2006},
    month = {01},
    issn = {1071-1023},
    doi = {10.1116/1.2140005},
    url = {https://doi.org/10.1116/1.2140005},
}

@article{DOLGOV2015708,
title = {Characterization of carbon contamination under ion and hot atom bombardment in a tin-plasma extreme ultraviolet light source},
journal = {Applied Surface Science},
volume = {353},
pages = {708-713},
year = {2015},
issn = {0169-4332},
doi = {https://doi.org/10.1016/j.apsusc.2015.06.079},
url = {https://www.sciencedirect.com/science/article/pii/S0169433215014233},
author = {A. Dolgov and D. Lopaev and C.J. Lee and E. Zoethout and V. Medvedev and O. Yakushev and F. Bijkerk}
}

@inproceedings{10.1117/12.934138,
author = {Yoshiharu Namba},
title = {{Specular Spectral Reflectance Of A1S1304 Stainless Steel At Near-Normal Incidence}},
volume = {0362},
booktitle = {Scattering in Optical Materials II},
editor = {Solomon Musikant},
organization = {International Society for Optics and Photonics},
publisher = {SPIE},
pages = {93 -- 103},
year = {1983},
doi = {10.1117/12.934138},
URL = {https://doi.org/10.1117/12.934138}
}

@article{NAMBA1980409,
title = {Surface Properties of Polished Stainless Steel},
journal = {CIRP Annals},
volume = {29},
number = {1},
pages = {409-412},
year = {1980},
issn = {0007-8506},
doi = {https://doi.org/10.1016/S0007-8506(07)61361-4},
url = {https://www.sciencedirect.com/science/article/pii/S0007850607613614},
author = {Yoshiharu Namba and Hideo Tsuwa}
}

@article{Yang:06,
author = {Minghong Yang and Alexandre Gatto and Norbert Kaiser},
journal = {Appl. Opt.},
number = {1},
pages = {178--183},
publisher = {Optica Publishing Group},
title = {Highly reflecting aluminum-protected optical coatings for the vacuum-ultraviolet spectral range},
volume = {45},
month = {Jan},
year = {2006},
url = {https://opg.optica.org/ao/abstract.cfm?URI=ao-45-1-178},
doi = {10.1364/AO.45.000178},
}

@article{Madden:63,
author = {R. P. Madden and L. R. Canfield and G. Hass},
journal = {J. Opt. Soc. Am.},
number = {5},
pages = {620--625},
publisher = {Optica Publishing Group},
title = {On the Vacuum-Ultraviolet Reflectance of Evaporated Aluminum before and during Oxidation$\ast$},
volume = {53},
month = {May},
year = {1963},
url = {https://opg.optica.org/abstract.cfm?URI=josa-53-5-620},
doi = {10.1364/JOSA.53.000620},
}

@article{DUNE:2021hwx,
    author = "Abud, A. Abed and others",
    collaboration = "DUNE",
    title = "{Design, construction and operation of the ProtoDUNE-SP Liquid Argon TPC}",
    eprint = "2108.01902",
    archivePrefix = "arXiv",
    primaryClass = "physics.ins-det",
    reportNumber = "FERMILAB-PUB-21-332-ND",
    doi = "10.1088/1748-0221/17/01/P01005",
    journal = "JINST",
    volume = "17",
    number = "01",
    pages = "P01005",
    year = "2022"
}
